\documentclass[12pt,a4papper]{article}
\usepackage{graphics}
\usepackage{graphicx}
\usepackage{amssymb}

\begin{document}

\title{The Pioneer anomaly as an effect of the dynamics of time}
\author{Antonio F. Ra\~nada\\Facultad de F\'{\i}sica, Universidad Complutense,\\
E-28040 Madrid, Spain}
\date{1 July 2004}
\maketitle

\begin{abstract}

 A model is presented in
which the Pioneer anomaly is not related to the motion of the
spaceship, but is a consequence of the acceleration of the
cosmological proper time $\tau$ with res\-pect to the coordinate
parametric time $t$, {\it i. e.} of the value of \mbox{${\rm
d}^2\tau/{\rm d}t^2\; (>0$)}. Assuming flat spatial sections and
that all the matter and energy is uniformly distributed throughout
the universe, it is shown that this inequality is an effect of the
background gravitational potential of the entire universe.
According to this model, the light speed, while being constant if
defined with respect to $\tau$ ({\it i. e.} as ${\rm d}\ell /{\rm
d} \tau$) as is required in general relativity, would suffer an
adiabatic secular acceleration, $a_\ell={\rm d}c/{\rm d}t >0$, if
defined in terms of $t$ ({\it i. e.} as ${\rm d}\ell /{\rm d} t$).
It turns out that such an adiabatic acceleration of light
(according to the second definition), and a small acceleration of
the Pioneer towards the Sun $a_{\rm P}$  could be mistaken the one
for the other, because they do have the same fingerprint: a blue
shift. However, this shift would be quite unrelated
 to any anomalous motion  of the Pioneer, being just
an observational effect of the acceleration of light with respect
to time $t$, in such a way that $a_{\rm P}= a_\ell/2$. A simple
estimation predicts $a_{\rm P}\simeq 5.2\times 10^{-10}\mbox{
m/s}$, just about $40\, \%$ smaller than the so-called Pioneer
acceleration, which would correspond to the blueshift. The Pioneer
anomaly turns out then to be an interesting case of the dynamics
of time, its explanation involving the interplay between the two
times $\tau$ and $t$.
  The view  presented here is the
relativistic version of a previous Newtonian model.

\end{abstract}



\vspace{4cm}

{\LARGE \bf Contents}

\bigskip

1. Introduction

\hspace{4mm} 1.1 The observation

\hspace{4mm} 1.2 The purpose of this paper

\hspace{4mm} 1.3 The dynamics of time and the two definitions of
light speed

\hspace{4mm} 1.4 Constant of light speed  and observed light speed

\hspace{4mm} 1.5 Equal fingerprints of two different phenomena

\hspace{4mm} 1.6 Plan of the paper

2. On the speed of light and the coordinate time

3. The background gravitational potential and the acceleration of
the

\hspace{4mm} non-proper light speed

4. Estimation of the adiabatic acceleration of the non-proper
speed of light

\hspace{4mm} 4.1 Time variation of the non-proper speed of light

\hspace{4mm} 4.2 Estimation of the non-proper accelerations of
light and of the acceleration of the clocks

5. The acceleration of cosmological time with respect to
coordinate time

\hspace{4mm} 5.1 A toy model

6. Summary and conclusions

\hspace{4mm}Acknowledgements

\hspace{4mm}References

\newpage


\section{Introduction}  \paragraph{1.1 The observation.} Anderson {\it et al} reported in 1998 the
observation of an anomalous acceleration $a_{\rm P}$ in the
Pioneer 10/11, Galileo and Ulysses spacecrafts, equal to  $a_{\rm
P}= (8.74\pm 1.33)\times 10^{-10}\mbox{m/s}^2$, constant and
directed towards the Sun \cite{And98}. More precisely, they
observed a Doppler blue shift in the radio signals from these
ships, which increases linearly in time, so that
\begin{equation}\dot{\nu}/\nu =2a_{\rm P}/c\, (=\mbox{const})\, ,\label{1.1.1}\end{equation}
where the factor 2 is because the Doppler effect refers to a
two-way signal. Obviously, its simplest interpretation is that the
ships were not following the predicted orbits but had an extra
unmodelled acceleration towards the Sun, as if our star pulled a
bit too much from them with a force independent of the distance.
 The effect is still unexplained and,
 intriguingly enough, does not show up in the planets \cite{And02}.
Anderson {\it et al} said in 1998: ``it is interesting to
speculate on the possibility that the origin of the anomalous
signal is new physics \cite{And98}, and in 2002: ``The veracity of
the signal is now undisputed, although the source of the anomaly,
some systematic or some not understood physics, is subject to
debate" \cite{And02b}.  For an interesting argument showing that
it may not necessarily be due to systematics, see \cite{Mbe04}.

\paragraph{1.2 The purpose of this paper.} In some previous papers
\cite{Ran03a,Ran03b,Ran03c}, a Newtonian model was presented that
proposes an explanation for the Pioneer riddle, showing that it
could be due to a universal and adiabatic acceleration of light,
which would have the same observational fingerprint as the
observed blue shift. That model is based on an application of the
fourth Heisenberg relation to the sea of virtual electron-positron
pairs which, combined with the expansion of the universe, would
produce a progressive decrease of the optical density of the
quantum vacuum, more precisely a progressive decrease of the
permittivity and permeability of empty space, the consequence
being a very small acceleration of light.  The purpose of this
paper is to propose a relativistic version of that Newtonian
model. It must be emphasized that a near Newtonian situation will
be always assumed in this paper, whenever the observed light speed
will be mentioned.

\paragraph{1.3 The dynamics of time and the two definitions of the light speed.}
A primordial element of this model is the dynamics of time, which
will be understood here as the functional relation $\tau =\tau
(t)$ between the cosmological proper time $\tau$ and the
parametric coordinate time $t$. It is important, therefore, to
emphasize that the light speed can be defined in general
relativity in two different ways, (i) with respect to the
cosmological proper time $\tau$, $c^*={\rm d}\ell /{\rm d} \tau \,
(=\mbox{constant})$, and (ii) with respect to the coordinate time
$t$, $c ={\rm d}\ell /{\rm d} t=c({\bf r}, t)$, where ${\rm
d}\ell$, ${\rm d} \tau$ and ${\rm d} t$ are elements of spatial
distance, proper time and coordinate time along a null geodesic,
respectively. These two speeds, both used in this paper, will be
denoted as $c^*$ and $c$, respectively, and will be called ``{\it
proper light speed}"  and ``{\it non-proper speed of light
speed}". The derivative of $c$ with respect to $t$ will be called
{\it non-proper acceleration of light}. The first is a universal
constant of nature, the second is not but, quite on the contrary,
depends generally on space and time $c=c({\bf r},t)$. Indeed, the
element of interval can be written (assuming for simplicity that
$g_{0i}=0$)
\begin{equation}
{\rm d}s^2=c^{*\, 2}{\rm d}\tau ^2-{\rm d}\ell ^2\,
,\label{1.3.1}\end{equation} with ${\rm d}\tau
=\sqrt{g_{00}}\,{\rm d}t$ and ${\rm d}\ell ^2=g_{ij}{\rm d}x^i{\rm
d}x^j$, so that $c^*$ is constant  and $c$ is
\begin{equation}
c=c({\bf r},t)=c^* \sqrt{g_{00}}\, .\label{1.3.2}\end{equation}
Near the Newtonian limit, $g_{00}\simeq 1+2\Phi /c^2$, at first
order, $\Phi$ being the gravitational potential, so that
\begin{equation}
c=c({\bf r},t)=c^*[1+\Phi ({\bf r},t)/c^2]\,
,\label{1.3.3}\end{equation} If $c_0$ is the value of $c$ at a
reference terrestrial laboratory $R$, {\it i.e.} $c_0=c^*(1+\Phi
_R/c_0^2)$,
\begin{equation}
c=c({\rm r},t)=c_0[1+\Phi ({\bf r},t)-\Phi _R/c_0^2]\,
,\label{1.3.4}\end{equation} The expression ``light speed" means
usually the constant $c^*$ which, being a universal constant, is
of the utmost importance. However, the non-proper light speed $c$
is also used in some interesting cases. For instance, in the study
of the bending of a light ray that grazes the Sun surface. It is
observed that the bending angle is $\phi =1.75^{\prime \prime}$.
Let $M$ and $R$ be the mass and radius of the Sun. The interval
around any star is given by the Schwarzschild metric, what implies
that $c= c(r)= c_\infty (1-\eta R/r)$, with $c_\infty =c(\infty)$
and $\eta =GM/c_\infty^2R\simeq 2.1 \times 10^{-6}$. Einstein gave
two formulae for this effect. The first (1907) is based only in
the equivalence principle and gives $\phi =2\eta = 0.875^{\prime
\prime}$, just one half of the observed effect. The second (1916),
in the frame of general relativity, gives the complete result
$\phi =4\eta = 1.75^{\prime \prime}$. The first one can be
obtained simply by considering the propagation of a wave light
with the previous value of the light speed, in other words as the
solution of the variational problem
\begin{equation}
\delta T=\delta \int _1^2{1+\eta R/r\over c_\infty}\, {\rm d}\ell
 =0\, ,\label{1.3.5}\end{equation} where ${\rm d}\ell={\rm
d}x^2+{\rm d}y^2+{\rm d}z^2$ is the euclidean line element, {\it
i.e.} as a consequence of the application of the Fermat principle
to the non-proper light speed $c$. The complete effect is obtained
by taking instead the non-euclidean spatial line element of the
Schwarzschild geometry. Note that this implies that the problem is
solved by assuming that the light propagates through with speed
$c= c_\infty (1-\eta R/r)$ (taking into account the Riemannian
character of the spatial metric). The same situation appears in
the study of the delay of the radar echoes from the planets.

It will be assumed in this paper, for simplicity, that all the
matter and energy in the universe are uniformly distributed. Since
the universe is expanding, it is clear that $c=c(t)$ and that
there is a functional relation between the two times $\tau =\tau
(t)$, determined by the background gravitational potential of all
the universe. All this means that
\begin{equation}
c(t)={{\rm d}\ell \over {\rm d}\tau}\,{{\rm d}\tau \over {\rm
d}t}= c^*\,{{\rm d}\tau \over {\rm d}t} \qquad \mbox{and}\qquad
{{\rm d}c(t)\over {\rm d}t}=c^*{{\rm d}^2\tau \over {\rm d}t^2}\,
.\label{1.0}\end{equation}  Equation (\ref{1.0}) states that {\it
the light speed $c$ increases if the proper time $\tau$
accelerates with respect to the coordinate time $t$}, its time
derivative being equal to the proper light speed $c^*$ times the
second derivative of $\tau$ with respect to $t$. The notation
$a_\ell =\dot{c}(t) ={\rm d}c(t)/ {\rm d}t $ will be used,
$a_\ell$ being termed the non-proper acceleration of light, or
just the acceleration of light for short when there is no risk of
confusion. It will be shown that $c(t)$ must increase as a
consequence of the interaction of light with the background
gravitational field due to all the matter and energy in e
universe, so that $a_\ell>0$. This will give an explanation for
the Pioneer anomaly.

It must be emphasized again that the proper speed of light $c^*$
is constant in this model, as it must happen in general
relativity. In fact, $c^*$ it is a universal constant of nature.
Expressions such as ``acceleration of light" or ``variation of
light speed" are always applied here to the light speed $c(t)$,
not to $c^*$. {\it In this sense, this model is not a variable
light speed theory} (of the class frequently known as VLS
theories), but a model inside general relativity. A standard of
velocity is still given by the constant $c^*$, which will allow to
measure the variation in time of the dimensional quantity $c(t)$
by means of measurements of the dimensionless value of $c(t)/c^*$.
The previous considerations are needed because the expression
``light speed" means several different things and we must
understand very carefully which one of its meanings we are using
(see \cite{Ell03} for an interesting discussion).

\paragraph{1.4 Constant of light speed  and observed light speed.}
 This paper presents a relativistic argument
which shows that the Pioneer anomaly could be a manifestation of
the adiabatic increase of the non-proper light speed $c$. But it
is often stated that it is impossible to measure an eventual
variation in time of $c$, because it is a dimensional constant.
This does not deny necessarily the existence of an acceleration of
light. However, since the possibility can not be excluded that the
standards of length or of time might be changing also, such a
measurement would be ambiguous or meaningless. According to this
view, widely shared by the metrologists, only non-dimensional
constants can be measured as functions of time, as would be the
case of the fine structure constant for instance.

However, this argument does not consider the fact that the
quantity known usually as ``the light speed" has two meanings: the
observed speed of light and a constant of physics that appears in
Einstein's equations  (in fact, it has more than two meanings but
these two are more important for the purpose of this work, see
\cite{Ell03}). It is usually assumed as a matter of fact that
these two faces are the same one, so that the values of the
universal constant (a theoretical concept) and of the observed
light speed (an empirical datum) are always identical. This seems
to be the case, but physics is an experimental science and this
must not be taken for granted necessarily. It can not be excluded
a priori that the two values could be different at large scales.
In that case, the observed light speed $c_{\rm obs}$ could be a
function of the constant light speed and the gravitational
potential $\Phi$ such as
\begin{equation}
c_{\rm obs}= c^*\times g\left(\Phi /c^{*\,2}\right)\, ,
\label{1.10}
\end{equation} at first order in potential, where $c^*$ is
the constant of nature  and $g$ is a dimensionless function. In
that case and if $\Phi$ depends on time, a time variation of the
light could be detected meaningfully by measuring the
dimensionless quantity $c_{\rm obs}/c^*=g(\Phi /c^{*\,2})$. Even
if the observed light speed changes in time, there would be still
a standard of speed provided by the constant $c^*$ (or any
function of $c^*$ which could be suitable).

\paragraph{1.5 Equal fingerprints of two different phenomena.} It turns out (and this must
be stressed) that an adiabatic increase of the coordinate time
speed of light $c$ (so that $a_\ell =\dot{c}>0$, see section 1.3)
has the same observational signature as a Doppler blue shift due
to the approaching motion of the source, as was shown in
\cite{Ran03b,Ran03c}. This can be understood easily, just by
taking the wave equation with variable speed for the field, for
instance $[\nabla ^2- c^{-2}(t)\partial _t^2]{\bf E}=0$, with
$c(t) =c_0+a_\ell \,t$ (the same result is obtained by using $\bf
H$ instead; note that {\it the time in this equation is clearly
the coordinate time}). At first order in $a_\ell$ the plane wave
solutions are ${\bf E}={\bf E}_0\exp\{{-2\pi i[z/\lambda -(\nu
_0+\dot{\nu}t/2)t]}\}$ with constant $\lambda$ and
\begin{equation}
\dot{\nu}/\nu _0 =a_\ell /c\, ,\label{1.4.1}\end{equation} which
must be compared with (\ref{1.1.1}) (the frequency is the time
derivative of the phase, {\it i.e.} ${\rm
d}[(\nu_0+\dot{\nu}t/2)t]/{\rm d}t=\nu _0+\dot{\nu}t=\nu
_0[1+a_\ell\, t/c]$). In other words, all the increase of the
speed is used to increase the frequency, the wavelength remaining
constant. This has two consequences: (i) an increase of non-proper
light speed $c$ with respect to the coordinate time $t$, $a_\ell
=2a_{\rm P}$, would produce an extra
 blue shift in radio signals,
linearly increasing in time and having the same observational
effect as the unmodelled acceleration equal to $a_{\rm P}$ towards
the Sun of the Pioneer and other spacecrafts, and (ii) the atomic
clocks would be accelerating, since their periods would be
decreasing. The blue shift would be due to the acceleration of
light, not to the motion of the ships which would be following
then and now the standard laws of gravitation without any extra
pull from the Sun. This means that the Pioneer anomaly would be
explained by finding a good reason for the coordinate time light
speed $c$ to increase.

\paragraph{1.6 Plan of the paper.}
 In
section 2,  the speed of light defined in terms of the coordinate
time $c(t)$ will be considered, with basis on some developments by
Einstein after proposing his equivalence principle and on the
ideas of electric permittivity and magnetic permeability of
gravitational origin. After that, it will be shown in section 3
that the background gravitational potential of all the universe
implies an acceleration of light with respect to the coordinate
time $t$, in the sense that $a_\ell=\dot{c}>0$ (see section 1.3).
Section 4 contains a simple estimation of $a_\ell$, which turns
out to be close, al least, to twice the Pioneer acceleration, the
value that would solve the riddle. Section 5 analyzes the dynamics
of time, {\it i. e.} the relation between $\tau$ and $t$, as the
basis for the explanation of the Pioneer anomaly: the speed of
light is constant, if defined with respect to the cosmological
proper time $\tau$, which in turn accelerates with respect to the
coordinate parametric time. The conclusions will be stated in
Section 6.

\section{On the speed of light and the coordinate time}
Einstein's papers of the period 1907-1912, often considered just
as matter for historians, are however of great interest for
physicists, in particular because of his discussions on the
variation of the light speed in a gravitational field \cite{Pai84,
Ron85}. Since he had not yet proposed his general relativity, with
the idea of proper time, he was obviously talking of the light
speed $c(t)={\rm d}\ell /{\rm d}t$ (called non-proper light speed
in section 1.3). In the last section of a review paper in 1907
\cite{Ein07}, he introduces his principle of equivalence in a
static situation, deducing from it that the light speed $c$ must
depend on the gravitational potential $\Phi ({\bf r})$, as well as
his first formula for the bending of a light ray grazing a star.
According to Pais, ``the study of Maxwell equations in accelerated
frames had taught him that the [observed] velocity of light is no
longer a universal constant in the presence of gravitational
fields" \cite{Pai84}.

In 1911 he takes anew the question in a paper entitled ``On the
influence of gravitation on the propagation of light"
\cite{Ein11}, where he uses again his principle of equivalence.
After a discussion on the synchronization of clocks, he concludes
that ``if we call the velocity of light at the origin of
coordinates $c_0$, where we take $\Phi =0$, then the velocity of
light at a place with gravitational potential $\Phi$ will be given
as
\begin{equation}
c=c_0\left(1+\Phi / c^2\right)\, ." \label{2.10} \end{equation} He
thus confirms, with a more detailed analysis, the conclusion of
his 1907 paper, stating also that eq. (\ref{2.10}) is a first
order approximation (although he does not tell which one, it is
obvious that he refers to situations near the Newtonian limit). It
can be written a bit more explicitly as
\begin{equation}
c({\bf r},t ) =c_0\left\{1+\Phi ({\bf r}, t)/c^2({\bf r},t)-\Phi
_{\rm R}/ c_0^2\right\}\, , \label{2.20}
\end{equation}
where $\Phi _{\rm R}$ is a reference potential, at present time in
a terrestrial laboratory $R$, where the observed light speed is
$c_0$, {\it i. e.} the constant that appears in the tables.
 It follows from equations (\ref{2.10})-(\ref{2.20}) that $c={\rm d}\ell
/{\rm d}t$ must depend on ${\bf r}$ and $t$, so that
 the deeper (more negative) is the potential, the smaller is
the light speed with respect to $t$ and conversely (according to
this Einstein formula, the light speed at the surface of the Sun,
would be about 2 ppm lower than here at Earth).

In two papers in 1912 \cite{Ein12a}, he  considers the light speed
as a field in spacetime $c({\bf r},t)$ and states ``a clock runs
faster the greater the $c$ of the location to which we bring it",
a statement quite similar to consequence (ii) at the introduction
of this work that the acceleration of light must cause or be
related to an acceleration of atomic clocks, to be discussed in
section 5. In a reply to a critical paper on relativity by M.
Abraham, Einstein states ``the constancy of the velocity of light
can be maintained only insofar as one restricts oneself to
spatio-temporal regions with constant gravitational potential.
This is where, in my opinion, the limit of the principle of the
constancy of the velocity of light --- thought not of the
principle of relativity
--- and therewith the limit of the validity of our current theory
of relativity, lies" \cite{Ein12b}. What Einstein says here is
that the principle of relativity is not the same thing as the
principle of constancy of light speed: the latter must not be
taken as a necessary consequence of the former. Furthermore, he
insists that the light velocity can depend on $\Phi$, as was the
case with his eq. (\ref{2.10}).

Two comments are important here: (i) as stated before, these
statements by Einstein refer clearly to the light speed defined
with respect to the coordinate time $c={\rm d}\ell /{\rm d}t$, as
explained in section 1.3, and (ii) they are still valid as first
order approximations in general relativity. As Einstein himself
found later, the light speed is constant and invariant if defined
with respect to proper time $c^*={\rm d}\ell /{\rm d}\tau$ (=
constant), but at that time he was mainly concerned with the
effect of a gravitational field on Maxwell equations, in which the
time derivatives are certainly with respect to the coordinate time
$t$. Let us consider now this question (I will follow now the well
known textbook {\it The Classical Theory of Fields} by Landau and
Lifshitz, reference \cite{Lan75}, section 90).

The electromagnetic tensor is defined in general relativity by
means of a vector field such that $ F_{\mu
\nu}=A_{\nu;\mu}-A_{\mu;\nu}=\partial _\mu A_\nu -\partial _\nu
A_\mu \, .$  The electromagnetic vectors $\bf E$, $\bf D$ and
antisymmetric tensors $B_{ij},\, H_{ij}$ are defined as follows
$$E_i=F_{0i}\, ,\quad B_{ij}=F_{ij}\, ,\quad
D^i=-\sqrt{g_{00}}\, F^{0i}\, ,\quad H^{ij}=\sqrt{g_{00}}\,
F^{ij}\, ,$$ the vectors $\bf B$, $\bf H$ being the dual to the
three-tensors $B_{ij}$ and $H_{ij}$, {\it i. e.} $B^i=-
e^{ijk}B_{jk}/(2\sqrt{\gamma})$, $H_i=-\sqrt{\gamma}\,
e_{ijk}H^{jk}/2$, where $\gamma =\det (\gamma _{ij})$, $\gamma
_{ij}= -g_{ij}$ being the three-dimensional metric tensor
(assuming for simplicity that $g_{0i}=0$). It follows  that
\begin{equation} {\bf D}={\bf E}/ \sqrt{g_{00}}\, , \quad {\bf
B}={\bf H}/ \sqrt{g_{00}}\, ,\label{1.1d}\end{equation} (see
\cite{Lan75}). If the space is empty, {\it i. e.} without free
charges or currents, the Maxwell equations can be written as
\begin{equation} \nabla \cdot {\bf B}=0\, , \quad \nabla \times
{\bf E}=-{1\over \sqrt{\gamma}}\, \partial _t\left(
\sqrt{\gamma}\, {\bf B}\right)\, .\label{1.1h}\end{equation}
\begin{equation}
\nabla \cdot {\bf D}=0\, ,\quad \nabla \times {\bf H}= {1\over
\sqrt{\gamma}}\,
\partial _t\,\left(\sqrt{\gamma}\,{\bf D}\right)\, .\label{1.1k}\end{equation} In a static
situation, these four equations have exactly the same form as in
special relativity, since the factors $\sqrt{\gamma}$ cancel.
However, eq. (\ref{1.1d}) implies that the relative permittivity
$\epsilon _{\rm r}$ and permeability $\mu _{\rm r}$ of empty space
are different from 1, their common value being $\epsilon _{\rm
r}=\mu _{\rm r}=(g_{00})^{-1/2}$. This is due to the geometry of
spacetime. Near the Newtonian limit one has $g_{00}=1+\Phi ({\bf
r})/c^2-\Phi _{\rm R}/c_0^2$, where $\Phi _{\rm R}$ is the
potential at a reference laboratory where $c=c_0$, so that the
empty space is like an inhomogeneous optical medium with $\epsilon
_{\rm r}({\bf r})=\mu _{\rm r} ({\bf r}) =1-[\Phi ({\bf
r})/c^2-\Phi _{\rm R}/c_0^2]$. Since $c=c_0/\sqrt{\epsilon _{\rm
r}\mu _{\rm r}}$, the light speed with respect to $t$ is being
given by Einstein eq. (\ref{2.20}) at first order.

In this work, we will be interested mainly in the case of a
potential depending only on time $\Phi (t)$, as is the case for
the background potential of all the universe in the approximation
that all its matter and energy are uniformly distributed. As will
be seen later, the time derivatives of the potential (over $c^2$)
 will be extremely small, of the order of the Hubble constant
$H_0=2.3\times 10^{-18}\mbox{ s}^{-1}$. The same applies,
therefore, to the time derivatives of $\gamma$. It is easy then to
deduce from eqs. (\ref{1.1h})-(\ref{1.1k}) that the
electromagnetic vector fields obey classical wave equations with
light speed given by Einstein's equation (\ref{2.20}), for
frequencies such that $\omega \gg H_0$, {\it i. e.} for any
practical purpose.

As a last remark to end this section, note that Einstein formula
(\ref{2.20}) can be written as
\begin{equation}
c({\bf r},t ) =c^*\left[1+\Phi ({\bf r}, t)/c^2({\bf
r},t)\right]\, , \label{201}
\end{equation}
where $c^*=c_0[1-\Phi _{\rm R}/c_0^2]$ is the light speed for zero
gravitational potential. It is also the light seed if defined with
respect to proper time, as will be discussed later.

It seems clear, therefore, that, according to
(\ref{2.10})-(\ref{2.20}), the light speed with respect to
coordinate time must be a function $c({\bf r},t)$, with both a
space and a time variation. The first depends on the distribution
of matter and energy. The second must be dominated by a secular
progressive increase, since the potential $\Phi$ produced by all
the matter and energy of the universe must be an increasing
function of time because of the universal expansion.

\section{The background gravitational potential and the acceleration of
the non-proper light speed. }  Let us take the element of interval
in weak gravity \cite{Mis73,Ber89,Wil93,Rin01}
\begin{equation} {\rm d} s^2= e^{2\Phi _{\rm loc}({\bf
r},t)/c^2}\,c^{*\, 2}\,{\rm d} t^2-{\rm d} \ell^2\approx (1+2\Phi
_{\rm loc} /c^2)\,c^{*\,2}\,{\rm d}t^2-{\rm d}\ell ^2\,
,\label{3.10}\end{equation} from which \begin{equation} c({\bf
r},t)=c^*[1+\Phi _{\rm loc}/c^2]\, ,\label{3.10.1}\end{equation}
at first order (compare with (\ref{201}), $\Phi _{\rm loc}({\bf
r})$ being here the (weak) gravitational potential of nearby
bodies, those that produce a non negligible acceleration $g({\bf
r})$ at the observation point (the Solar System and the Galaxy,
for observers near Earth).  One has thus for the light speed at a
generic point $P$ along a ray, $ c(P)=c(R)[1+(\Phi _{\rm
loc}(P)-\Phi _{\rm loc}(R))/c^2],$ $R$ being a reference point.

Note that (\ref{3.10.1}) is the same as Einstein formula. It
implies that the difference  between the non-proper light speed
and the $\tau$-light speed is an effect of the gravitational
potential since they are equal if $\Phi =0$. The local variations
of $\Phi$ due to the inhomogeneities of the distribution of matter
give local and small variations of $c({\bf r}, t)$, while its
cosmological variation causes a secular increase which is seen as
the blueshift, observed as the Pioneer effect.

It will be important in the following to know what kind of time is
the variable $t$ in (\ref{3.10}). It is certainly a coordinate
time, but this is not enough for our purpose, in view of the
freedom to make changes of the spacetime coordinates. Since this
element of interval is good near the classical limit, this $t$ is
a Newtonian time. In fact, the increment of the interval of a
particle world line  between the points $P_1$ and $P_2$ is
$$\Delta s=\int_{P_1}^{P_2}{\rm d}s =\int
_{t_1}^{t_2}c\left(1+{2\Phi _{\rm loc}\over c^2}-{v^2\over
c^2}\right)^{1/2}{\rm d}t\, ,$$ which, at first order in the
potential, is equal to
\begin{equation}
\Delta s = c(t_2-t_1)-{1\over c}\int_{t_1}^{t_2} \left({v^2\over
2}-\Phi _{\rm loc}\right){\rm d}t\label{3.15}\end{equation} As is
seen, the condition for $\Delta s$ to be stationary is equivalent
to the Hamiltonian principle for a particle in the potential $\Phi
_{\rm loc}$. This shows that, for weak gravity, the time $t$ in
(\ref{3.10}) is a Newtonian parametric time.

But we are not only submitted  to the potential nearby bodies.
Quite on the contrary, there is a background gravitational
potential due to all the matter and and energy in the visible
universe, which is assumed here to be uniformly distributed over
flat surfaces $t=\mbox{ constant}$, so that the background
gravitational potential is a function of time $\Phi _{\rm
all}(t)$. That potential must be uniform and depend on time
because of the expansion. We must therefore rewrite the interval
(\ref{3.10}) as
\begin{equation}
{\rm d} s^2= c^{*\, 2}\,{\rm d} \tau^2-{\rm d} \ell^2\, ,\quad
\mbox{with }\, {\rm d}\tau =[1+\Psi (t)]{\rm d}t\,
,\label{3.16}\end{equation} at first order, where the
dimensionless potential $\Psi (t)=\Phi _{\rm all}(t)/c^2(t)$ the
background potential of all the matter and energy. From now on,
the notation $\Psi =\Phi /c^2$ will be used.

It follows that
\begin{equation}
c=c(t)=c^*[1+\Psi (t)] =c_0[1+\Psi (t)-\Psi (t_0)]\,
,\label{3.16.1}\end{equation} where $t_0$ is the age of the
universe, {\it i.e.} the present time, and $c_0=c(t_0)=c^*[1+\Psi
(t_0)]$, at first order, is the present time value of the
non-proper light speed at a terrestrial reference laboratory ({\it
i.e.} the value in the tables). Note that this equation gives the
variation of $c(t)$ near $t_0$ at first order. Taking now the $t$
derivative of (\ref{3.16.1}) at time $t_0$, one has
$\dot{c}(t_0)=c(t_0)\dot{\Psi}(t_0)$. The same argument can be
applied to the expression for $c(t)$ near any other fixed time
$\tilde{t}$, what implies that $c(t)=c(\tilde{t})\exp[\Psi
(t)-\Psi (\tilde{t})]$ $\forall t$. In particular, taking
$\tilde{t}=t_0$, one finds
\begin{equation}c(t)=c_0\,e^{[\Psi (t)-\Psi (t_0)]}\, .\label{3.30}\end{equation}
The shape of the function $c(t)$ defined by (\ref{3.30}) does not
change if the reference time ($t_0$ or $\tilde{t}$) is changed
because $c(t_1)e^{-\Psi (t_1)}=c(t_2)e^{-\Psi (t_2)}$. The
interval can be written then as
\begin{equation}
{\rm d}s^2=e^{2[\Psi (t)-\Psi (t_0)]}c_0^2{\rm d}t^2-{\rm d}\ell
^2=c^{*\, 2}{\rm d}\tau ^2-{\rm d}\ell ^2\, ,\quad  \mbox{with }
{\rm d}\tau =e^{\Psi (t)}{\rm d}t\, .\label{300}\end{equation}
Note that $\tau$ is a well defined cosmological proper time. We
see here that $\tau$ accelerates with respect to $t$, its second
derivative being not nil. This will be considered later with more
detail.

 The interval (\ref{3.16}) is
correct near the Newtonian limit, even if $\Psi (t)$ is not small,
since $\nabla \Psi =0$. Indeed, if such a potential is included in
(\ref{3.15}), the Lagrangian will increase in a function of time,
which does not have any effect on the equations of motion of the
particle. This fact will be used in section 4. In any case, $\Psi$
being space independent, it can be absorbed in a redefinition of
the time. As a last remark in this section, note that the
gravitational potential $\Phi$ has been used even in cases in
which its gradient can be neglected. For instance, by Ahluwalia
and coworkers to study quantum aspects of gravity, eventual
violations of the equivalence principle or gravitationally induced
neutrino-oscillation phases, see \cite{Ahl97,Ahl01}.

To summarize this section, the background gravitational potential
of all the universe induces an acceleration of proper time with
respect to coordinate time. Since the $\tau$-light speed is
constant, the non-proper light speed necessarily accelerates.

\section{Estimation of the adiabatic acceleration of the non-proper speed of
light}
\paragraph{4.1 Time variation of the non-proper speed of light.}  The effect of the expansion on
the potential at a spacetime point will be considered now, an
estimate being made later in section 4.2 of the non-proper
acceleration of light $a_\ell=\dot{c}(t_0)$ (\ref{1.0}). The
potential of all the universe at the terrestrial laboratory $R$
can be written, with good approximation, as $\Phi _{\rm all}=\Phi
_{\rm loc}(R)+\Phi_{\rm av}(t)$. The first term $\Phi _{\rm
loc}(R)$ is the part due to the  local inhomogeneities, {\it i.
e.} the nearby bodies (the Solar System and the Milky Way). It is
constant in time since these objects are not expanding. The second
$\Phi_{\rm av}(t)$ is the space averaged potential due to all the
mass and energy in the universe (except for the nearby bodies),
assuming that they are uniformly distributed. Contrary to the
first, it depends on time because of the expansion. The former has
a non vanishing gradient but is small, the latter is space
independent,
 but time dependent and much larger. The
 value of $\Phi _{\rm
loc}/c_0^2$ at $R$ is the sum of the effects of the Earth, the Sun
and the Milky Way, which are about $-7\times 10^{-10}$, $-10^{-8}$
and $-6\times 10^{-7}$, respectively, certainly with much smaller
absolute values than $\Phi _{\rm av}(t_0)$, which of the order of
$-10^{-1}$ as will be seen below.

Taking the time derivative of eq. (\ref{3.16.1}) or (\ref{3.30}),
the (observed) light speed near present time $t_0$ is equal to
\begin{equation} c(t)=c_0[1+a_{\rm t} (t-t_0)]=c_0+a_{\rm \ell}(t-t_0),
\label{4.30}\end{equation} the quantity $a_{\rm t}$ and the light
acceleration $a_\ell$ being
\begin{equation}   a_{\rm t} = \dot{\Psi} (t_0)\, ,\quad \quad
 a_\ell =a_{\rm t}c_0=\dot{\Psi}(t_0)c_0\, ,\label{4.40}
\end{equation}
Since the background gravitational potential of all the universe
$\Phi _{\rm av} (t)$ is increasing because of the expansion (the
galaxies are separating and their interaction potential
increasing) eqs. (\ref{3.16.1})-(\ref{4.40}) show that a time
increase of the non-proper speed of light  $c(t)$  must be
expected also. In this sense, there is an acceleration of light
$a_\ell =\dot{c}$ (see section 1.3).
 Indeed, the arguments
leading to (\ref{4.30}) are clear and compelling. Unfortunately, a
rigorous calculation of the quantities $a_\ell$ and of $a_{\rm
t}$, which would take into account all the eventual effects, is
not easy. However, a simple, approximate and phenomenological
estimation will be performed now. It is claimed that it is
sensible and meaningful in spite of its simplicity (it may be
convenient to stress again that all this is compatible with the
constancy of the $\tau$-light speed $c^*={\rm d}\ell/{\rm
d}\tau$.)

\paragraph{4.2 Estimation of the non-proper accelerations of
light and of the acceleration of the clocks.}
 The inverse time $a_{\rm t}$ was in fact introduced by Anderson
 {\it et al} in reference \cite{And98}, as the ``clock acceleration", through the relation
 ``$a_{\rm P}=a_{\rm t}c$". It will be shown later
in section 5 that it is indeed the acceleration of the atomic
clocks.

In the following a simple crude estimate of the values of the
 non-proper acceleration of light (section 1.3)
$a_\ell=\dot{c}(t_0)$ and the clocks acceleration will be made,
taking (\ref{3.16.1}) as starting point. Although it involves
approximations and simplifications, it shows the main ideas of the
model and gives an convincing representation of the phenomenon.
Let $\Omega _M,\,\Omega _\Lambda$ be the corresponding present
time relative densities of matter (ordinary plus dark) and dark
energy corresponding to the cosmological constant $\Lambda$. We
take a universe with $k=0$, $\Omega _M=0.27$, $\Omega _\Lambda
=0.73$ and Hubble parameter $H_0=71\mbox{ km}\cdot
\mbox{s}^{-1}\cdot \mbox{Mpc}^{-1}= 2.3\times 10^{-18}\mbox{
s}^{-1}$.
 In
order to determine the average potential $\Phi _{\rm av}(t)$, let
$\Phi _0(t_0)$ be the gravitational potential produced by the
critical mass density distributed up to the present radius of the
visible universe $R_{\rm U}(t_0)=c_0/H_0=4,200\mbox{ Mpc}$; taking
into account the kinetic energy of the galaxies as a source of
gravity, one has $\Phi _0(t_0)/c_0^2=-\int _0^{c_0/H_0}
c_0^{-2}G\rho _{\rm cr}4\pi r{\rm d}r/\sqrt{1-(H_0r/c_0)^2}=4\pi
G\rho_{\rm cr}/H_0^2=-0.75$. It must be emphasized that, although
this value of the potential might seem to be too large for this
approximation to apply, there is no problem in fact since it is
space independent and its time derivative is extremely small, as
explained at the end of section 3. Its effect will be to
accelerate adiabatically the proper time with respect to the
coordinate time, as will be seen later.

Because, in this model, the light speed was smaller in the past,
the radius of the visible universe is a function of time $R_{\rm
U}(t)$ that can be written as $R_{\rm U}(t) =R_{\rm U}(t_0)\,[\int
_0^tc(t){\rm d}t\, / \int _0^{t_0}c(t){\rm d}t]\, .$ It turns out
then that $\Phi _0(t)=4\pi G\rho_{\rm cr}R_{\rm U}^2(t)/c_0^2$.
Consequently, one has $\dot{\Phi}_0(t_0)=\Phi _0(t_0)\times
(2c_0/R_U)$. Note that $\Phi _0(t)\rightarrow 0$ when
$t\rightarrow 0$.

The present time average potential is then $\Phi _{\rm
av}(t_0)=\Phi _0(\Omega _M-2\Omega _\Lambda$). Because of the
expansion of the universe, the gravitational potentials due to
matter and dark energy equivalent to the cosmological constant
vary in time as the inverse of the scale factor $R(t)$ and as its
square $R^2(t)$, respectively (with $R(t)=\left({\Omega _M/ \Omega
_\Lambda }\right)^{1/3}\sinh ^{2/3}\left[{(3\Lambda )^{1/2}t/
2}\right]$ for this model universe). This implies that the average
background gravitational potential is given as
\begin{equation}
\Phi _{\rm av}(t)=\Phi _0(t_0)\left[{\int _0^tc(t){\rm d}t\over
\int _0^{t_0}c(t){\rm d}t}\right]^2\left[{\Omega _M\over
R(t)}-2\Omega _\Lambda R^2(t)\right]\, .\label{4.50}\end{equation}

 From  eqs. (\ref{3.16.1})-(\ref{4.50}) and a bit of simple algebra, the
inverse time $a_{\rm t}$ can be expressed as
$$a_{\rm t}={[1-3\Omega _\Lambda]\dot{\Phi} _0/c_0^2-[1+3\Omega
_\Lambda]H_0\Phi _0/c_0^2\over 1+2(1-3\Omega _\Lambda )\Phi
_0/c_0^2}=H_0{(1-9\Omega _\Lambda )\Phi _0/c_0^2\over
1+2(1-3\Omega \Lambda )\Phi _0/c_0^2}.$$ Introducing in this
equation the values of $\Omega _M,\,\Omega _\Lambda, \Phi _0$ and
$\dot{\Phi}_0$,  the clock acceleration $a_{\rm t}$ and the light
acceleration $a_\ell$ are shown to take the values
\begin{equation} a_{\rm t}\simeq 1.5\,H_0\, , \quad a_\ell =a_{\rm t}
c_0\simeq 10.4 \times 10^{-10}\mbox{ m/s}^2\,
.\label{4.80}\end{equation}

 Since the anomaly would be explained if $a_{\rm P}=a_\ell /2$
 and the observed value is $a_{\rm P}= (8.74\pm
1.33)\times 10^{-10}\mbox{m/s}^2$, the predicted Pioneer
acceleration only about 40 $\%$ off the observation. This is
encouraging for such a simple estimate, but note that the main
purpose of this work is less to get a precise fit of $a_{\rm P}$
than to build an understanding of the phenomenon. It could not be
otherwise, given the simplicity of the calculation. What matters
is that the model predicts the existence of an adiabatic
non-proper acceleration of light, {\it i.e.} an increase of the
non-proper light speed $c={\rm d}\ell/{\rm d}t$ implying a blue
shift with a value close, at least, to the one observed in the
Pioneer 10 and the spaceships. This is what would give an
explanation of the riddle.

It must be underlined that such an acceleration of light is a
simple consequence of Einstein formula (\ref{2.20}), taking into
account the expansion of the universe, since the potential
$\Phi_{\rm av}$ is increasing because of the expansion, so that
$c(t)$ must increase also (see section 1.3). This increase, in
turn, produces a blue shift as shown in section 1.4, with the same
observational signature as an extra attraction from the Sun. It
could be argued, however, that the gravitational redshift would
cancel the increase in $c$ since the frequency decreases when
photons climb up along an increasing potential. However, the well
known expression for the gravitational redshift $\Delta \nu /\nu_0
=-\Delta \Phi /c_0^2$ is not valid here, since it assumes static
situations which is not the case in an expanding universe. To end
this section, it can be said that Rosales has proposed an
explanation for the Pioneer anomaly which, although different to
the one considered here since he argues that it is a manifestation
of the Berry phase, coincides in attributing the effect to the
expansion of the universe \cite{Ros04}.

\section{The acceleration of cosmological time with respect to coordinate time}
The main result of this paper is the predicted value of the
Pioneer acceleration in section 4.2. This was made locally in
time, in terms of some data at present time $t_0$, as the
background gravitational potential of all the universe and its
derivative with respect to the coordinate time $t$, assuming a
near Newtonian situation. This section contains the development of
some ideas based in a cosmological and global consideration of
time.

As explained in section 2, Einstein asserted in  1912 that clocks
run faster the higher is $c$ (page 104 of the first paper of
\cite{Ein12a}; remember, his definition of the light speed was
$c(t)={\rm d}\ell /{\rm d}t$). This means that the higher is $c$
the larger is the quotient $\Delta \tau /\Delta t$, {\it i.e.} an
interval of proper time over the corresponding interval of
coordinate time. He was thinking then in clocks at different space
points, but the same can be said clearly about clocks at different
times: if he had known the universal expansion (unsuspected
however at that time) he could have added ``and, besides, clocks
run faster as time goes on". The (cautious) claim of this and
previous work is that the Pioneer acceleration is an effect of the
acceleration of light with respect to the coordinate time $t$ (see
section 1.3). These two statements are equivalent. Indeed, if the
observed blue shift is the consequence of the acceleration of
light, the basic units of time ({\it i. e.} the periods of
electromagnetic waves) would be decreasing (see section 1.4), the
atomic clocks going faster, and conversely. Otherwise stated, the
Pioneer effect could be a manifestation of the acceleration of
time, as measured by atomic clocks. In other words, of the
acceleration of proper time with respect to coordinate time.

 In (\ref{3.16.1}) he proper time $\tau$ was shown to be given as
\begin{equation} {\rm d}\tau =\sqrt{g_{00}}\; {\rm d}t=e^{\Psi (t)}{\rm
d}t\, ,\quad  \tau (t)=\int _{t_i}^te^{\Psi(t)}\,{\rm d}t \,
,\label{5.20}
\end{equation} $t_i$ being an arbitrary initial time.
 As stated before, the time $\tau$ accelerates with respect to
$t$, the acceleration being obviously equal to
\begin{equation}{{\rm d}^2\tau \over {\rm
d}t^2}=\dot{\Psi}(t)e^{\Psi (t)}= a_{\rm t}e^{\Psi (t)}\, ,
\label{5.46}\end{equation}because of the universal expansion.

However, a warning is necessary: although $\tau$ is well defined
and is the cosmological proper time (assuming a uniform
distribution of mass and energy), it is not the time used in
measurements with atomic clocks at Earth. Indeed, since the
frequencies increase at the same rate as the light speed, the
basic units of these clocks vary on time as $1/c(t)$. This means
that, near present time $t_0$, the time of the atomic clocks, say
$\tau_{\rm at}(t;t_0)$, verifies
\begin{equation} {\rm d}\tau _{\rm at}(t;t_0)=e^{[\Psi
(t)-\Psi(t_0)]}{\rm d}t\, ,\quad \left.{{\rm d}^2\tau _{\rm
at}(t;t_0)\over {\rm d}t^2}\right|_{t_0}= \dot {\Psi}(t_0)=a_{\rm
t}\, .\label{5.50}\end{equation} As is seen, $a_{\rm t}$ is really
the acceleration of the atomic clocks, as already stated by
Anderson et al \cite{And98}.  The speed of light, if measured or
calculated with respect to this time, is constant and equal to
\begin{equation} c[\tau _{\rm at}(t;t_0)]={{\rm d}\ell\over {\rm
d}\tau _{\rm at}(t;t_0)} =c_0\,,\quad \mbox{while }\;\;c(\tau)
={{\rm d}\ell\over {\rm d}\tau} =c^* \,.\label{5.55}\end{equation}
In other words, the measurements of the light speed with such
atomic clocks gives the value of $c(t)$ not of $c^*$. However,
$\tau _{\rm at}(t;t_0)$ is not a good universal time since its
definition depends on $t_0$ (the observation time); nevertheless
it is good near $t_0$.
  It verifies ${\rm d}\tau _{\rm at}(t;t_0)=e^{- \Psi(t_0)}{\rm d}\tau (t)$, and
 ${\rm d}^2\tau _{\rm at}/{\rm d}\tau ^2=0$. This means that
these two times, $\tau _{\rm at}$ and $\tau$, do not accelerate
with respect to one another, their time intervals being just
proportional, the proportionality constant being $e^{-\Psi
(t_0)}$. Furthermore ${\rm d}\tau _{\rm at} (t_0;t_0)={\rm d}t$,
what means that, at exactly the time $t_0$, the corresponding
intervals of the times $\tau _{\rm at}$ and $t$ are equal. This
clarifies the meaning of $\tau _{\rm at}(t;t_0)$: {\it it is a
redefinition of the cosmological proper time (by means of a
multiplicative factor depending on $t_0$), to ensure that the
basic units of $\tau _{\rm at}$ and $t$ are equal at $t_0$ }({\it
i. e.} now).

Otherwise stated, the introduction of the factor $e^{-\Psi (t_0)}$
makes sure that the atomic clocks which are used tick at time
$t_0$ at the same rate (if more exactly) as the mechanical
classical clocks that measure parametric time $t$. This explains
why a measure with the clocks with time $\tau _{\rm at}(t;t_0)$
gives the value $c(t_0)$, not $c^*$. Since they are made to tick
at the same rate as the coordinate time (as the barycenter dynamic
time of the solar system, for instance) they give the same value
for shorts time intervals near $t_0$. On the other hand a true
measurement with atomic clocks would employ clocks that ticked at
the same rate as the mechanical clocks at some initial time in the
past without redefining their intervals, so that their ``seconds",
would be different now. The result of such a measurement would be
the constant proper speed of light $c^*$.

Let us summarize which are the values of the light speed near
present time $t_0$, as defined or measured with respect to the
three times $t$, $\tau$ and $\tau _{\rm at}(t;t_0)$, eqs
(\ref{4.30}) and (\ref{5.55}),
\begin{equation}c(t)=c_0[1+a_{\rm t} (t-t_0)]\, ,\qquad c(\tau )=c^*\, ,\qquad
c[\tau _{\rm at}(t;t_0)]=  c_0\, . \label{5.60}\end{equation}
 As is seen, if $c$ is defined:
(i) with respect to the coordinate parametric time $t$, it
increases linearly in $t$; (ii) with respect to the proper
cosmological time $\tau (t)$, it is constant and equal to $c^*$;
and (iii) with respect to the time of the atomic clocks
synchronized with $t$ at time $t_0$, $\tau _{\rm at}(t;t_0)$, it
is constant and equal to $c_0$.

Assume, now, that we change $t_0$ to $t_0+\Delta t_0$, $\Delta t_0
\, (\lll t_0)$ being the time difference between two measurements
of the light speed with atomic clocks ({\it e. g.} 1 year or 20
years). Taking into account that $\Psi (t_0+\Delta t_0)=\Psi
(t_0)+a_{\rm t}\Delta t_0$ and instead of (\ref{5.55}), we will
have for the light speed, at first order, if measured with respect
to time $\tau _{\rm at}(t;t_0+\Delta t_0)$,
\begin{eqnarray}c[\tau _{\rm at}(t;t_0+\Delta t_0)]&=&{{\rm d}\ell\over {\rm d}\tau _{\rm at}(t;t_0+\Delta t_0)}
={1\over [1+\Psi (t)-\Psi (t_0)][1-a_{\rm t}\Delta t_0]}{{\rm
d}\ell \over {\rm d}t}\nonumber\\ &=& {{\rm d}\ell \over {\rm
d}t}\,(1+a_{\rm t}\Delta t_0)=c_0(1+a_{\rm t} \Delta t_0)\,
 ,\label{5.70}
\end{eqnarray}
instead of (\ref{5.55}).  This means that, in a measurement of the
light speed with atomic clocks at times $t_0$ and $t_0+\Delta t_0$
(if synchronized with coordinate time at the time of the
measurement), one would find an acceleration of light with respect
to $t$ equal to $a_\ell =a_{\rm t}c_0$, as in (\ref{4.40}). This
is the conclusion of this model: that the Pioneer anomaly is an
effect of the dynamics of time \cite{Tie02,Bar98}.

\paragraph{5.1 A toy model.} This paper is about what happens now or
in an interval around present time $t_0$, its main result being
the prediction of an adiabatic acceleration of light $a_\ell$ with
respect to the coordinate time, eqs. (\ref{4.10})-(\ref{4.30}),
which would have the same observational signature as the Pioneer
effect. However, this needs some clarification. What was shown is
that {\it the observed light speed is constant if measured with
respect to the cosmological proper time $\tau$, while $\tau$
accelerates with respect to the coordinate time $t$. As a
consequence, light accelerates with respect to coordinate time
$t$, so that $\dot{c}(t)=a_\ell>0$}. Moreover, a very simple
approximate estimation gives for $a_\ell$ a value just a $40\,\%$
smaller than twice the Pioneer acceleration, the value that would
explain exactly the effect (see section 1.3). Given the simplicity
of the calculation, this seems encouraging.

 It is tempting to explore as well the model towards the past, as a
method to understand better the problem. The result of this
exploration will be called now ``a toy model" to emphasize that it
is not taken necessarily as a rigorous theory. In spite of that,
it may be useful to understand better the main ideas involved,
from an intuitive point of view. To do that it suffices to take
$t_i=0$ in  (\ref{5.20}), so that
\begin{equation}  \tau (t)=\int
_0^te^{\Psi(t)}{\rm d}t,\qquad c=c_0\,e^{[\Psi
(t)-\Psi(t_0)]}=c^*e^{\Psi(t)} \, ,\label{5.80}\end{equation} with
$\Psi (t)=\Phi_ {\rm av}(t)/c^2(t)$, $\Phi _{\rm av}(t)$ being
given in (\ref{4.50}).
 Let us play with this toy model, in order to build an intuitive picture of the
 phenomenon. It must be underlined that, if the light speed were smaller in the past,
 it can be admitted that the universe is also smaller than what is now admitted.  As an
 approximation, we will define the present value of the potential,
 introduced in section 4.2 as
 $$\Phi _0(t_0)/c_0^2=-\int _0^{R_{\rm U}}
c_0^{-2}G\rho _{\rm cr}4\pi r{\rm d}r/\sqrt{1-(r/R_{\rm U})^2}\,
,$$ and change the value of $R_{\rm U}$. The corresponding Hubble
parameter is  $H_0= c_0/R_{\rm U}$.

\begin{figure}[h]
\begin{center}
\scalebox{1}{\includegraphics{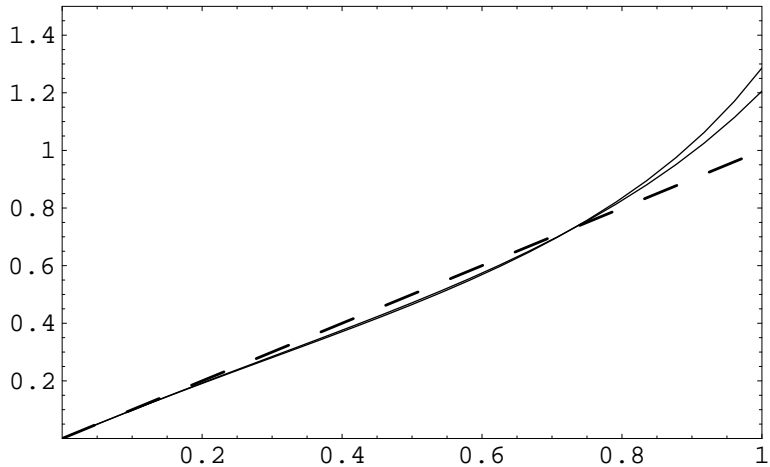}}
\end{center}
\caption{Cosmological proper time $\tau$ (\ref{5.80}) in the toy
model (solid lines)
  and coordinate parametric time $t$ (dashed line) versus
parametric time, in units of the age of the universe $t_0$. The
lower (resp. upper) solid line at $t=t_0$ corresponds to $R_{\rm
U}(t_0)=3,000\mbox{ Mpc}$ (resp. $R_{\rm U}(t_0)=3,400\mbox{
Mpc}$)(explanation in the text). } \label{fig1} \label{Figure1}
\end{figure}

The most interesting results are summarized in two
  figures (the calculations are only approximate, no attempt having been made
  to refine the precision of the  numerical values).
In Figure \ref{Figure1}, the cosmological proper time $\tau$
(\ref{5.80}) for $R_{\rm U}(t_0)=3,000\mbox{ Mpc}$  and $R_{\rm
U}(t_0)=3,400\mbox{ Mpc}$  and the coordinate time $t$ are plotted
versus the coordinate time in units of the age of the universe.
The line of the coordinate $t$ is obviously a straight line. On
the other hand, the acceleration of $\tau$ with respect to $t$ is
seen in the curvature of the solid lines. Indeed, the other two
curves verify ${\rm d}^2\tau /{\rm d}t^2 <0$ for $t<t_{\rm
c}\simeq 0.34\,t_0$, and  ${\rm d}^2\tau /{\rm d}t^2
>0$ for $t>t_{\rm c}$.
 Before $t_{\rm c}$, $\tau$ decelerates
with respect to $t$, but it accelerates afterwards. The reason for
that behavior is that  the negative potential due to matter is
dominant at the beginning, while near present time the positive
potential due to the cosmological constant is more important (with
the assumed value of $\Omega _\Lambda$). The latter begins to take
over the former at time $t_{\rm c}$ because
 the potential $\Psi (t)$ had a minimum then.

It happens that $\tau (t_0)\simeq 1.20\,t_0$ (resp. $1.27\, t_0$)
for 3,000 Mpc (resp. 3,400 Mpc). In other words, the age of the
universe, as measured by $\tau$, would be about $1.20\,t_0$ (resp.
$1.27\, t_0$), while it is just $t_0$ in terms of $t$. This would
be due to the adiabatic acceleration $a_{\rm t}e^{\Psi (t)}$ of
$\tau$ with respect to $t$. Note that the difference between
$\tau$ and $t$ was small until recent times.

\begin{figure}[h]
\begin{center}
\scalebox{1}{\includegraphics{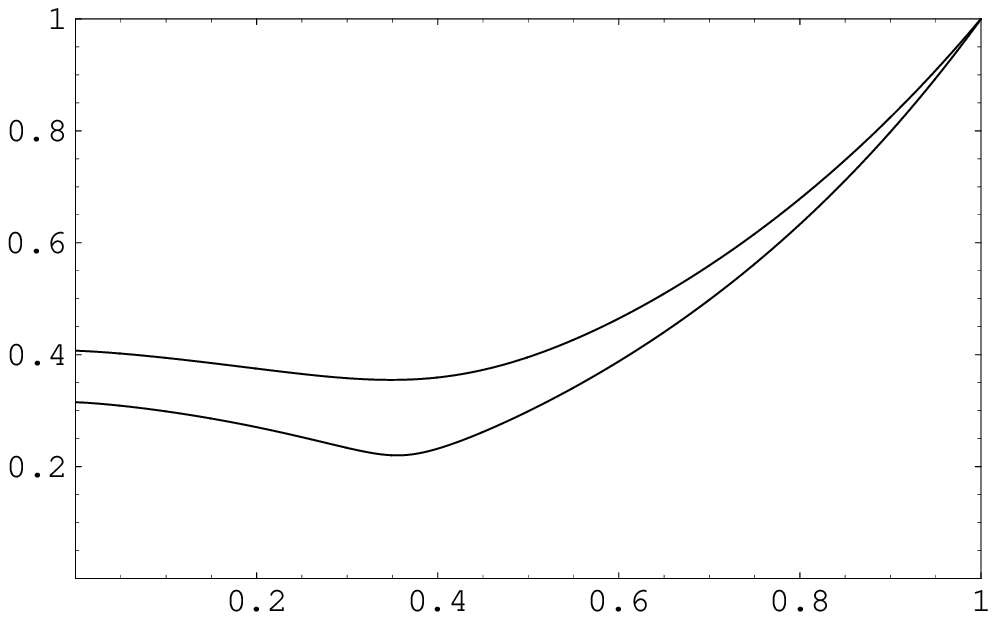}}
\end{center}
\caption{Non-proper light speed (\ref{5.80}) in the toy model, in
units of $c_0$, as function of the coordinate parametric time, in
units of the age of the universe $t_0$ (explanation in the text).}
\label{fig2} \label{Figure2}
\end{figure}

In Figure 2, the non-proper light speed (\ref{5.80}) is plotted
against $t$ for $R_{\rm U}(t_0)=3,000\mbox{ Mpc}$ (upper line) and
$R_{\rm U}(t_0)=3,400\mbox{ Mpc}$ (lower line). As is shown, it
decreases from $t=0$ until $t\simeq 0.34\,t_0$, where it has a
minimum, increasing thereafter. It does not change much, however,
before $t\simeq 0.5\, t_0$. Note that their values for $t=0$ turns
out to be $c^*\simeq 0.42\, c_0$ and $c^*\simeq 0.31\, c_0$,
respectively. Also that (i) at $t=\tau=0$, both definitions of
light speed give the same value $c(t=0)=c(\tau =0)=c^*$, and (ii)
$c(t)$ increases for $t>0$, while $c(\tau )$ remains constant, so
that $c>c(\tau )$, their difference increasing after the
beginning.

It is seen in both figures that at about $t=0.35\,t_0$ both the
cosmological proper time $\tau$ and the non-proper speed of light
$c$ begin to accelerate with respect to $t$, this effect being
more clear after about $t=0.5\,t_0$. This is the transit from the
slowing-down to the speeding-up, due to the effect of the dark
energy (represented here by the cosmological constant).

There is a singularity at about $R_{\rm U}\simeq 3,400\mbox{
Mpc}$. When that value is approached from below, the minimum of
$c(t)$ at $t\simeq 0.35\, t_0$ becomes more accused, the second
derivative ${\rm d}^2c/{\rm d}t^2$ increasing, until the equation
(\ref{5.80}) fails to have a real solution for $c$ around that
value of $t$. This happens approximately when the Schwarzschild
radius of the total mass of a distribution with the critical
density inside a sphere of radius $R_{\rm U}$ coincides with
$R_{\rm U}$, although this is perhaps just a numerical
coincidence.

\section{Summary and conclusions} In spite of its simplifications
and approximations, the model presented here gives a promising
explanation of the Pioneer anomaly, as a consequence of the
interplay between the cosmological and proper time $\tau$ and the
coordinate and parametric time $t$. All through this work, it is
assumed that (i) all the matter and energy in the universe are
uniformly distributed and (ii) the near Newtonian approximation is
acceptable. The idea of background gravitational potential of all
the universe $\Phi _{\rm av}(t)$ is important here. The
conclusions of this paper are the following:

1. {\it The light speed can be defined in two ways: (i) with
respect to the proper time $\tau$, $c^*={\rm d}\ell /{\rm d}\tau$,
and (ii) with respect to the coordinate time $t$, $c(t)={\rm
d}\ell /{\rm d}t$} (see section 1.3). With the first definition it
is a universal constant $c^*$, with the second it is a function
$c({\bf r},t)$. They have been called here ``proper speed of
light" and ``non-proper speed of light", respectively. Assuming a
uniform distribution for all the matter and energy of the
universe, the second definition gives a time dependent  non-proper
speed of light $c(t)$, the variation of which is dominated by a
secular adiabatic increase, due to the progressive augmentation of
the background gravitational potential $\Phi _{\rm av}(t)$ of all
the universe as the galaxies separate. In this sense it is
sensible to speak of an acceleration of light, even if the proper
speed of light is constant. The present value of that acceleration
would be $a_\ell =\dot{c}(t_0)=a_{\rm t}c_0$, $a_{\rm t}$ being
the coordinate time derivative of the background potential of all
the universe $\Psi (t)=\Phi _{\rm av}(t)/c^2(t)$ and $c_0$ the
value of the light speed in the tables. A simple estimate predicts
the value $a_\ell=10.4\times 10^{-10}\mbox{ m/s}^2$, what would
mean an increase of $c(t)$  of about 3.3 cm/s per year. Note,
however, that {\it this model is not a theory with variable light
speed}, since the proper speed of light is constant.

 2. {\it Such an adiabatic acceleration has
the same observational signature, an extra blue shift, as an
acceleration towards the Sun of a radio source such as the
Pioneer}. The agreement would be quantitatively exact if
$a_\ell\simeq 2a_{\rm P}$, {\it i. e.} if the light acceleration
were equal to twice the Pioneer acceleration (section 1.5).
However, $a_{\rm P}$ would be quite unrelated to any unmodelled
motion, although it could be easily interpreted as an anomalous
acceleration, even if the ship were following the exact trajectory
predicted by the current theory of gravitation.

3. The phenomenon here reported is due to the relation between the
cosmological proper time $\tau$ and the coordinate time $t$. That
relation is given as $\tau =\int _0^te^{\Psi (t)}{\rm d}t$, so
that so that ${\rm d}^2\tau/{\rm d}t^2= \dot{\Psi}\,e^\Psi$
(overdot means derivative with respect to $t$)  (section 3 and 4).
  In other
words, $\tau$ accelerates with respect to $t$, since ${\rm
d}^2\tau / {\rm d}t^2=\dot{\Psi}(t)e^{\Psi (t)}>0$. Consequently,
while the light  has constant speed with respect to $\tau$, it
accelerates with respect to $t$.
 As all this indicates, the
Pioneer phenomenon is a very interesting case of the dynamics of
time \cite{Tie02,Bar98}.

An important aspect of the problem refers to the detection of the
blueshift. The time used in the Solar System is the barycenter
dynamical time, which is originally a Newtonian parametric time,
even if it is measured with atomic clocks in order to increase the
precision. The blueshift can indeed be measured, as was the case,
by using detectors with circuits calibrated in a macroscopic way.
This is an indication in favor of this work. In any case, there is
an interesting metrological problem.

To summarize: the conclusion of this paper is that the anomalous
acceleration of the Pioneer 10/11 and the other two spacecrafts
could be only apparent, not real, just an effect of the dynamics
of time ({\it i. e.} the relation between he cosmological proper
time and the coordinate parametric time) that shows up in the
acceleration of light, if defined as the $t$ derivative of $c={\rm
d}\ell /{\rm d}t$ (the proper light speed $c^*={\rm d}\ell /{\rm
d}\tau$ is constant as required in general relativity). This would
cause the observed blue shift, which would be quite unrelated,
however, to the motion of the space ships. Indeed, they would have
followed the standard trajectories, as predicted by current
gravitation theory. All this must be studied by the experts who
know the details of the motion of the spacecrafts and of the
metrological procedures involved in the observation. This should
be done, since some of the consequences of the interplay between
the two times $t$ and $\tau$ could be unexpected and surprising at
the cosmological level.

{\bf Acknowledgements.} I am indebted to A. Tiemblo for
enlightening  discussions on the dynamics of time, to J. L.
Sebasti\'an for information on microwaves and to G. \'Alvarez
Galindo, C. Aroca, A. I. G\'omez de Castro, E. L\'opez and J. L.
Rosales for discussions or encouragement.

\end{document}